\documentclass[a4paper,11pt]{article}

\usepackage{contribution}

\makeatletter
\@ifundefined{c@affiliation}%
{\newcounter{affiliation}}{}%
\makeatother
\newcommand{\affiliation}[2][]{\setcounter{affiliation}{#2}%
  \ensuremath{{^{\alph{affiliation}}}\text{#1}}}
\newcommand{\weblink}[2][]{%
    \ifthenelse{\equal{#1}{}}%
    {\textnormal{\url{#2}}}%
    {\textnormal{\href{#2}{#1}}}%
}

\def\beq{\begin{equation}}
\def\eeq#1{\label{#1}\end{equation}}
\def\eeqn{\end{equation}}

\def\beqa{\begin{eqnarray}}
\def\eeqa#1{\label{#1}\end{eqnarray}}
\def\eeqan{\end{eqnarray}}

\let\bar=\overbar

\def\etal{{\it et al.}}

\def\Dslash{\not{\hbox{\kern-4pt $D$}}}
\def\dslash{\not{\hbox{\kern-2pt $\del$}}}

\def\msb{{\bar{\ssstyle M \kern -1pt S}}}

\newcommand{\contribution}[7][]{%
  \clearpage
  \thispagestyle{plain}
  \ifthenelse{\equal{#1}{}}
  {\hypersetup{pdftitle={#2}}}
  {\hypersetup{pdftitle={#1}}}
  \hypersetup{pdfauthor={{#3} {#4}}}
  {\centering\normalfont\LARGE\bfseries\sffamily #2 \par\nobreak}
  \lhead{}
  \chead{%
    \textit{\footnotesize XIV International Conference on Hadron Spectroscopy
      (\weblink[\textit{hadron2011}]{http://www.hadron2011.de}), 13-17 June 2011, Munich, Germany}%
  }
  \rhead{}
  \bigskip
  \begin{center}
    {#3} {#4}\ifthenelse{\equal{#6}{}}{}{\footnote{\weblink[#6]{mailto:#6}}}
    \ifthenelse{\equal{#7}{}}{}{#7} \\
    \textit{#5}
  \end{center}
  \bigskip
}

\renewcommand{\abstract}[1]{%
  \begin{center}
    \begin{minipage}{0.85\textwidth}
      \begin{footnotesize}
        #1
      \end{footnotesize}
    \end{minipage}
  \end{center}
  \bigskip
}

\begin{document}

%
%
%
%
%
{  

%
%

%
\contribution[Effective $Q-Q$ Interaction in Heavy Baryons]
{Effective Quark-Quark Interaction\\ in Heavy Baryons}
{Joseph P.}{Day}  
{\affiliation[Theoretical Physics, Institute of Physics,
  University of Graz,
  A--8010 Graz, Austria]{1} }
{joseph.day@uni-graz.at}
{\!\!$^,\affiliation{1}$, Ki-Seok Choi\affiliation{1}, and Willibald Plessas\affiliation{1}}
%
%

\abstract{%
  We report results from a study of heavy-baryon spectroscopy within a relativistic constituent-quark model, whose hyperfine interaction is based on Goldstone-boson-exchange dynamics. While for light-flavor constituent quarks it is now commonly accepted that the effective quark-quark interaction is (predominantly) furnished by Goldstone-boson exchange -- due to spontaneous chiral-symmetry breaking of quantum chromodynamics at low energies -- there is currently still much speculation about the light-heavy and heavy-heavy quark-quark interactions. With the increasing amount of experimental data on heavy-baryon spectroscopy these issues might soon be settled. Here, we show, how the relativistic constituent-quark model with Goldstone-boson-exchange hyperfine interactions can be extended to charm and bottom baryons. It is found that the same model that has previously been successful in reproducing the light and strange baryon spectra is also in line with the existing phenomenological data on heavy-baryon spectroscopy. An analogous model with one-gluon-exchange hyperfine interactions for light-heavy flavors does not achieve a similarly good performance.
}
%

\section{Framework}
We view hadrons as relativistic bound states of constituent quarks Q. Baryons are
are thus considered as $\{QQQ\}$ systems. Even if heavy-flavor quarks are involved, it is
mandatory to work in a relativistic framework in order to prevent pathologies and/or severe
shortcomings. Our relativistic constituent-quark model (RCQM) is based on a relativistically
invariant mass operator $\hat{M}=\hat{M}_{\text{free}} + \hat{M}_{\text{int}}$ that includes
the $Q$-$Q$ interaction according to the Bakamijian-Thomas construction~\cite{Bakamjian:1953}.
In the rest frame of the baryon, ${\vec P}=\sum^3_i \vec k^2_i=0$, the free and interacting
parts of the mass operator thus read
\begin{equation}
\hat M_{\text{free}}=\sum^3_{i=1}\sqrt{\hat m^2_i+\hat {\vec k}_i^2} \;, \hspace{10mm}
\hat M_{\text{int}}=\sum^3_{i<j}\hat V_{ij}=
\sum^3_{i<j}(\hat V^{\text{conf}}_{ij}+\hat V^{\text{hf}}_{ij}) \;,
\end{equation}  
where the $\vec{k}_i$ correspond to the three-momenta of the individual quarks with rest masses
$m_i$ and the mutual $Q$-$Q$ potentials $\hat V_{ij}$ are composed of confinement and
hyperfine interactions.
By employing such a mass operator $\hat M^2= \hat P^\mu \hat P_\mu$\,, with baryon
four-momentum $\hat P_\mu=(\hat H, \hat P_1, \hat P_2, \hat P_3)$, one satisfies the
Poincar\'{e} algebra involving all ten generators
$\{\hat H, \hat P_i, \hat J_i, \hat K_i\}$ of time and space translations, spatial rotations
as well as Lorentz boosts, respectively:
	\begin{equation}
     \begin{array}{lll}
	[\hat P_i,\hat P_j]=0~, & [\hat J_i~,\hat H]=0~, & [\hat P_i,\hat H]=0~,  \\ \nonumber
	[\hat K_i,\hat H]=i\hat P_i~, & [\hat J_i,\hat J_j]=i\epsilon_{ijk}\hat J_k~, &
    [\hat J_i,\hat K_j]=i\epsilon_{ijk}\hat K_k~, \\ \nonumber 
	[\hat J_i,\hat P_j]=i\epsilon_{ijk}\hat P_k~, &
	[\hat K_i,\hat K_j]=-i\epsilon_{ijk}\hat J_k~, &
	\mbox{$[\hat K_i,\hat P_j]=i\delta_{ij}\hat H~.$} \nonumber
	\end{array}
	\end{equation}
The solution of the eigenvalue equation of the mass operator $\hat M$ yields the
relativistically invariant mass spectra as well as the baryon eigenstates (in the standard
rest frame)~\cite{Melde:2008yr}.

\section{Goldstone-Boson Exchange in the $SU(3)_F$ Sector}
In order to incorporate the property of spontaneous breaking of chiral symmetry of low-energy
QCD, the Graz group has proposed a RCQM with Goldstone-boson-exchange (GBE) hyperfine
interactions between constituent quarks~\cite{Glozman:1998ag}. In addition to a linearly rising
confinement potential, as following from lattice QCD, the model comes with a spin-spin
hyperfine interaction of the form
\begin{equation}
\label{potsu3}
V^{\text{hf}}_{ij} = \bigg[V_{\pi}\sum_{a=1}^3 \lambda^a_i \lambda^a_j + V_{K}\sum_{a=4}^7 \lambda^a_i\lambda^a_j  + V_{\eta} \lambda^8_i \lambda^8_j +V_{\eta'}\lambda^0_i \lambda^0_j \bigg] \vec{ \sigma}_i \cdot \vec{\sigma}_j \,,
\end{equation}
where the $\lambda^a_i$ are the $SU(3)_F$ Gell-Mann matrices and $\vec{ \sigma}_i$ the $SU(2)_S$
Pauli matrices of quark $i$. The GBE is cast into the exchange of pseudoscalar mesons
$\pi$, $K$, and $\eta$, including the singlet $\eta'$ due to the $U(1)_A$ anomaly. The
corresponding meson-exchange potentials as a function of the relative $Q$-$Q$ distance
$\vec r_{ij}$ read
\begin{equation}
\label{rgbe}
V_{\gamma}(\vec r_{ij})= \frac{g^2_{\gamma}}{2 \pi} \frac{1}{12 m_i m_j} \bigg[ \mu^2_{\gamma} \frac{e^{-\mu_{\gamma} r_{ij}}}{r_{ij}} -\Lambda_{\gamma}^2 \frac{e^{-\Lambda_{\gamma}r_{ij}}}{r_{ij}}\bigg] \,, \hspace{5mm} \gamma = \pi, K, \eta, \eta' \,,
\end{equation}
where $g_{\gamma}$ is the meson-quark coupling constant, $\mu_{\gamma}$ the mass of the
exchanged meson, and $\Lambda_{\gamma}$ an adjusted cut-off parameter. As immediately evident,
the GBE RCQM produces a spin- and flavor-dependent hyperfine interaction. It has hitherto
been quite successful not only in describing the spectroscopy of all baryons with flavors
$u$, $d$, and $s$ in a unified framework but also in a number of baryon
reactions~\cite{Plessas:2010pk}. 

\section{Extension of the GBE RCQM to Heavy Baryons}

We have investigated, if the same model can be extended to include also baryons with flavors
$c$ and $b$ in a consistent manner. Thus we have generalized the hyperfine interaction of
eq.~(\ref{potsu3}) to $SU(5)_F$ in the following manner
\begin{eqnarray}
\label{potsu5}
V^{\text{hf}}_{ij} &=& \bigg[ V_{\pi}\sum_{a=1}^3 \lambda^a_i \lambda^a_j + V_{K}\sum_{a=4}^7 \lambda^a_i\lambda^a_j  + V_{\eta_8} \lambda^8_i \lambda^8_j + V_{\eta_0}\lambda^0_i \lambda^0_j \\ \nonumber
&&+V_{D}\sum_{a=9}^{12} \lambda^a_i \lambda^a_j + V_{D_s}\sum_{a=13}^{14} \lambda^a_i \lambda^a_j 
+ V_{\eta_{15}} \lambda^{15}_i \lambda^{15}_j \\ \nonumber
&&+V_{B}\sum_{a=16}^{19} \lambda^a_i \lambda^a_j + V_{B_s}\sum_{a=20}^{21} \lambda^a_i \lambda^a_j +V_{B_c}\sum_{a=22}^{23} \lambda^a_i \lambda^a_j + V_{\eta_{24}} \lambda^{24}_i \lambda^{24}_j\bigg] \vec\sigma_i \cdot \vec\sigma_j~,
\end{eqnarray}
where the various meson-exchange potentials are assumed in the same form as in eq.~(\ref{rgbe}).
The complete parametrization of the $SU(5)_F$ GBE RCQM can be found in ref.~\cite{day}.

Obviously, this new construction also influences the $Q$-$Q$ interaction in the $SU(3)_F$
sector (specifically through the diagonal elements in the flavor matrices $\lambda_{15}$ as
well as $\lambda_{24}$). Therefore our first concern is to maintain the good performance
regarding the baryon spectroscopy with $u$, $d$, and $s$ flavors. In Fig.~\ref{fig:strange}
we give a selective comparison of the $SU(3)_F$ and $SU(5)_F$ models vis-\`a-vis the
experimental data.

\begin{figure}[t]
  \begin{center}
   \includegraphics[width=0.7\textwidth]{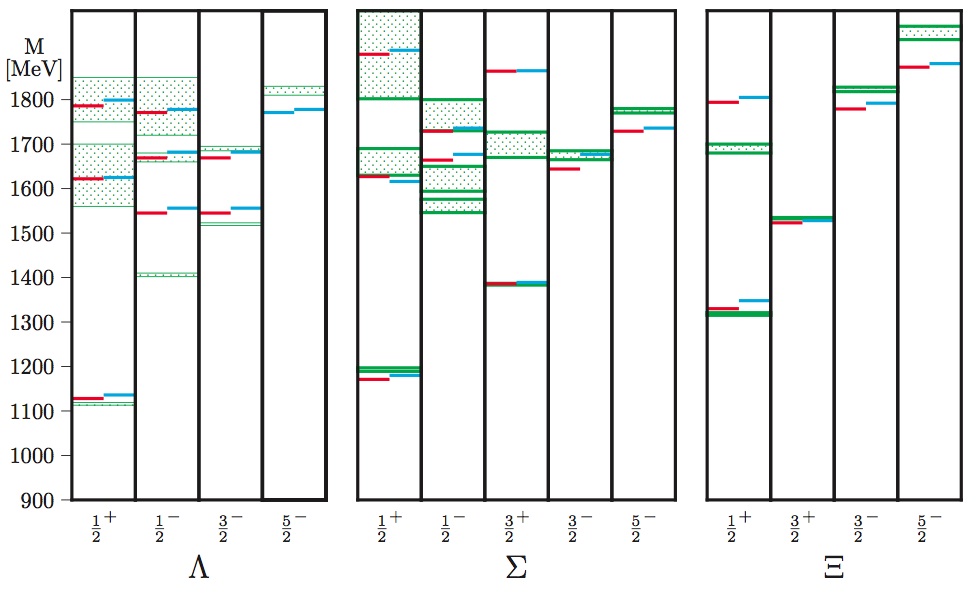}
   \caption{Comparison of $SU(5)_F$ (left/red levels) and $SU(3)_F$ (right/blue levels)
   hyperfine interactions in the GBE RCQM for the examples of the $\Lambda$, $\Sigma$, and $\Xi$
   baryons with definite spin and parity $J^P$ in each column. The green boxes represent the
   experimental data with their uncertainties as 
   given by the PDG~\cite{pdg}.}
   \label{fig:strange}
  \end{center}
\end{figure}

It is found that the quality of light and strange baryon spectroscopy is comparable for
both the $SU(3)_F$ and $SU(5)_F$ models. In the latter, some of the levels, such as the
$J^P=\frac{1}{2}^+$ $\Lambda$ ground state, are even closer to experiment.
A similar agreement occurs for the $N$, $\Delta$, and $\Omega$ states. In particular,
due to its specific flavor dependence, also the $SU(5)_F$ model produces the correct
orderings of positive- and negative-parity excitations simultaneously in the $N$ and
$\Lambda$ spectra.


\begin{figure}[t]
  \begin{center}
    \includegraphics[height=5.5cm]{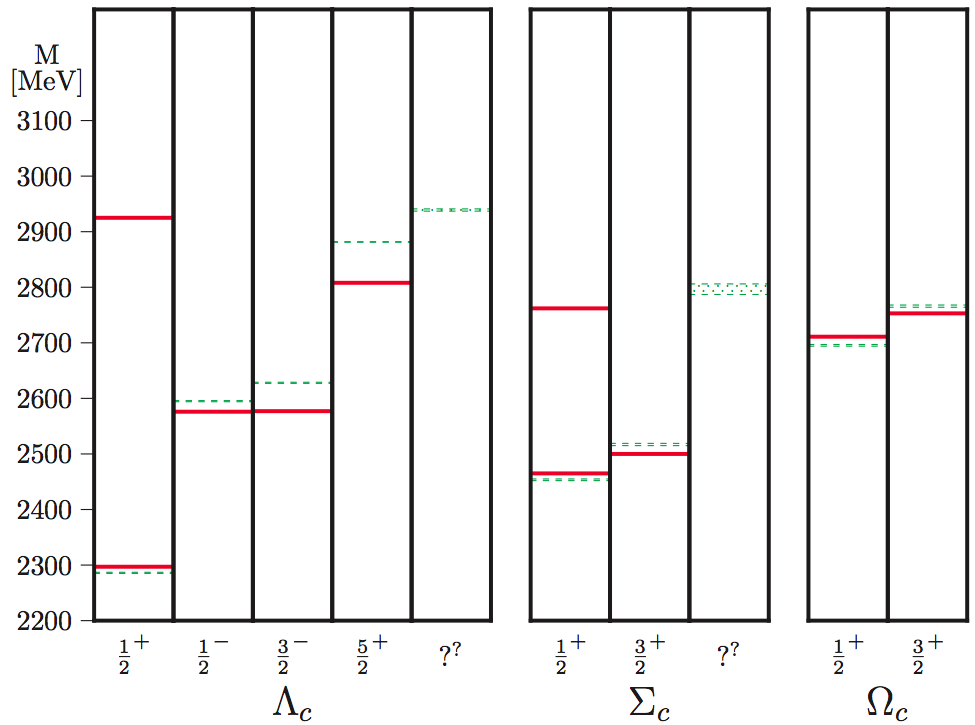}
    \includegraphics[height=5.5cm]{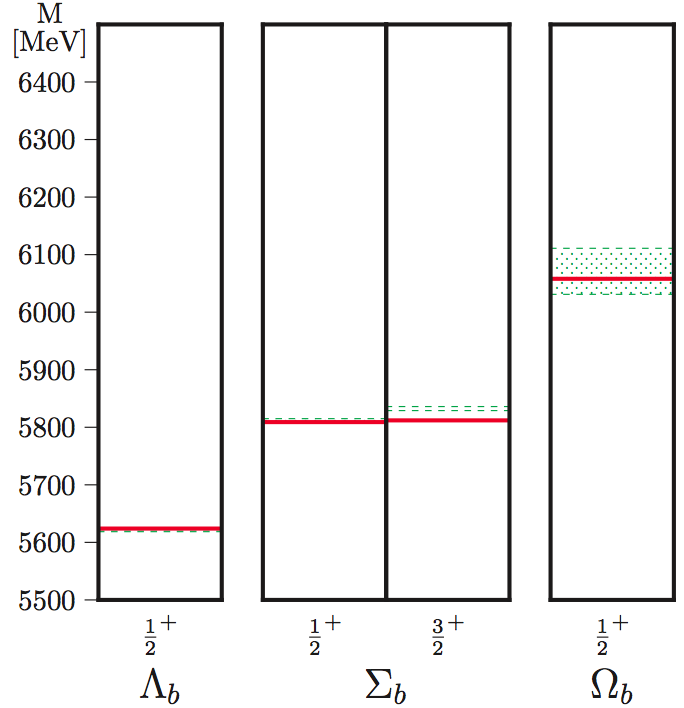}
    \caption{Heavy-baryon spectra as produced by the $SU(5)_F$ GBE RCQM (solid/red levels)
    in comparison to experimental data with their uncertainties, as quoted by the
    PDG~\cite{pdg} (dotted/green levels resp. boxes).}
    \label{fig:heavy}
  \end{center}
\end{figure}

In Fig.~\ref{fig:heavy} we show the results of the extended $SU(5)_F$ GBE RCQM for some
heavy-baryon spectra. It appears that all levels can be reproduced satisfactorily. The
same is true for the other cases not shown here because of space limitations. 


At this stage we have succeeded in constructing a universal RCQM based on GBE dynamics
that works in the whole baryon spectroscopy within $SU(5)_F$. For heavy baryons, additional
experimental data are highly desirable in order to better determine the model parameters and
put the theory to more stringent tests. Similarly, applications to describing reactions
with heavy baryons (e.g., form factors and decays) will be most interesting to study.  

\subsection*{Acknowledgment}

\vspace{-3mm}
This work was supported by the Austrian Science Fund, FWF, through the Doctoral
Program on {\it Hadrons in Vacuum, Nuclei, and Stars} (FWF DK W1203-N08).

%

}  


\end{document}